\documentclass[prb,twocolumn,superscriptaddress,showpacs,preprintnumbers,amsmath,amssymb]{revtex4}% Physical Review B

\usepackage{graphicx}% Include figure files
\usepackage{dcolumn}% Align table columns on decimal point
\usepackage{bm}% bold math
\usepackage{natbib}
\usepackage{amsmath}
\usepackage{amssymb}
\usepackage{float}
\usepackage{datetime}
\usepackage[normalem]{ulem}

\usepackage[usenames,dvipsnames]{xcolor}
\newcommand{\rodcolor}{\color{black}}

\begin{document}
\author{Stefano Roddaro}
\email{s.roddaro@sns.it}
\affiliation{NEST, Scuola Normale Superiore and Istituto Nanoscienze-CNR, Piazza S. Silvestro 12, I-56127 Pisa, Italy}
\affiliation{Istituto Officina dei Materiali -- CNR, Basovizza S.S. 14 km 163.5, I-34149 Trieste, Italy}
\author{Daniele Ercolani}
\affiliation{NEST, Scuola Normale Superiore and Istituto Nanoscienze-CNR, Piazza S. Silvestro 12, I-56127 Pisa, Italy}
\author{Mian Akif Safeen}
\affiliation{NEST, Scuola Normale Superiore and Istituto Nanoscienze-CNR, Piazza S. Silvestro 12, I-56127 Pisa, Italy}
\author{Francesco Rossella}
\affiliation{NEST, Scuola Normale Superiore and Istituto Nanoscienze-CNR, Piazza S. Silvestro 12, I-56127 Pisa, Italy}
\author{Vincenzo Piazza}
\affiliation{Center for Nanotechnology Innovation @NEST, Istituto Italiano di Tecnologia, Piazza San Silvestro 12, 56127 Pisa, Italy}
\author{Francesco Giazotto}
\affiliation{NEST, Scuola Normale Superiore and Istituto Nanoscienze-CNR, Piazza S. Silvestro 12, I-56127 Pisa, Italy}
\author{Lucia Sorba}
\affiliation{NEST, Scuola Normale Superiore and Istituto Nanoscienze-CNR, Piazza S. Silvestro 12, I-56127 Pisa, Italy}
\author{Fabio Beltram}
\affiliation{NEST, Scuola Normale Superiore and Istituto Nanoscienze-CNR, Piazza S. Silvestro 12, I-56127 Pisa, Italy}

\title{\rodcolor Large thermal biasing of individual gated nanostructures}

\begin{abstract}

We demonstrate {\rodcolor a novel nanoheating scheme that yields} very large and uniform temperature gradients {\rodcolor up to about $1\,{\rm K}$ every $100\,{\rm nm}$, in an architecture which is compatible with the field-effect control of the nanostructure under test. The temperature gradients demonstrated largely exceed those typically obtainable with standard resistive heaters fabricated on top of the oxide layer. The nanoheating platform is demonstrated in the specific case of a short-nanowire device.}

\end{abstract}

\pacs{72.20.Pa, 81.07.Gf, 85.30.Tv}

% 72.20.Pa	  Thermoelectric effects - in semiconductors and insulators
% 81.07.Gf	  Materials - nanoscale materials - nanowires
% 85.30.Tv		Transistors - field effect

\maketitle

In the past decade much effort was directed to the investigation of the thermoelectric (TE) properties of innovative materials. Such a revival of TE science was largely driven by the interest in solid-state energy converters~\cite{DiSalvo99,Bell08,Subramanian06,BookTE} and by the development of {\rodcolor novel advanced materials~\cite{Snyder08} and, in particular,} nanomaterials~\cite{Majumdar04,Vineis10,Shi12}. Indeed, the achievement of an efficient and cost-effective TE technology depends on the optimization of a set of interdependent material parameters of the active element: the Seebeck coefficient $S$ and the heat and charge conductivities $\kappa$ and $\sigma$. Recent developments in nanoscience yielded new strategies for the design of novel and more efficient nanomaterials in which the strong interdependency between $S$, $\kappa$ and $\sigma$ can be made less stringent~\cite{Boukai08,Venkatasubramanian01,Poudel08,Heremans08}. Despite the host of available theoretical predictions~\cite{Heremans08,Dresselhaus07,Zhang11,Shi10,Humphrey05}, however, the optimization of the TE behavior of nanostructured materials still remains an open and actively investigated problem~\cite{Wu13, Svensson13}, in particular for what concerns the influence of electron quantum states engineering on the power factor $\sigma S^2$. This led to the development of a number of experimental arrangements designed to impose a controllable thermal bias over micrometric or even submicrometric active elements and to measure how this affects charge transport in the device. Differently from macroscopic active elements, nanoscale TE materials also allow the investigation of thermal effects in devices where field-effect can be used to control carrier density~\cite{Wu13,Hochbaum09} {\rodcolor or even quantum states energetics~\cite{Roddaro11,Romeo12} and coupling~\cite{Hoffmann09}.} While this may not be a directly scalable strategy in view of applications, it is particularly useful for what concerns the fundamental investigation of the impact of doping {\rodcolor - a key parameter -} on TE performance. Various examples of microheating systems were reported in the literature. These include (i) suspended SiN$_x$ microheaters, which enable a precise estimate of the $\kappa$ of individual nanostructures, but also pose non-trivial technical challenges~\cite{Moore2011,Zhou2011} and do not allow the field-effect control of the nanostructure behavior; (ii) resistive heaters fabricated on top of standard Si/SiO$_2$ substrates, which are instead typically used to estimate $S$ and allow also the field-effect control of carrier density~\cite{Hoffmann09,Hochbaum09,Small03,Small04,Tian12,Roddaro13}.

\begin{figure}[h!]
\begin{center}
\includegraphics[width=0.48\textwidth]{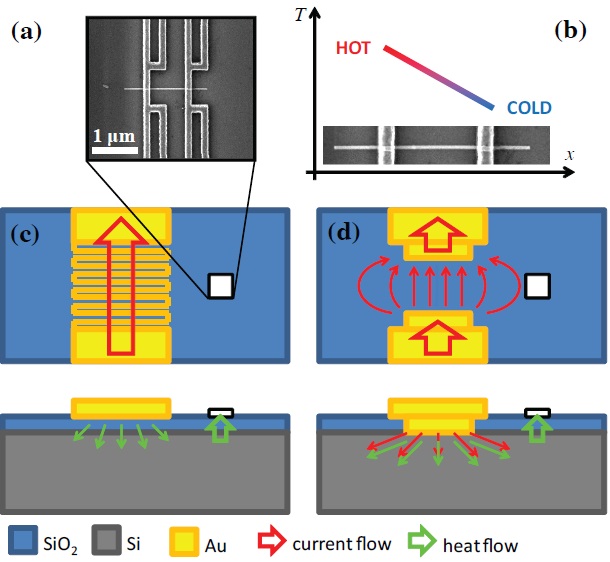}
\caption{\rodcolor The measurement of field-effect dependence of thermoelectric effects in single nanostructures (a nanowire device is visible in panel (a)) requires the application of a strong thermal gradient (panel (b)). A standard approach consists in the fabrication of a top heating element (panel (c)). An alternative ``buried'' architecture exploiting current flows into the bulk is visible in panel (d).}
\label{fig:Cartoon}
\end{center}
\end{figure}

Here we demonstrate an innovative buried-heater (BH) scheme based on current diffusion in the conductive bulk of a SiO$_2$/Si substrate. This scheme is different from the more standard one of ``top'' heaters (THs) relying on resistive elements microfabricated on top of the oxide layer. We shall show that our architecture yields very large and uniform thermal gradients {\rodcolor easily} exceeding $5\,{\rm K/\mu m}$ {\rodcolor and up to about $10\,{\rm K/\mu m}$}, far beyond typical values reported in the literature for THs. In addition, similarly to the case of TH architectures, our scheme allows the control of the nanostructure behavior by field effect. A sketch of the two alternative TH and BH schemes is visible in Fig. 1. The TH scheme relies on the diffusion of heat from a metallic resistive element through the oxide, into the substrate and thus below the nanostructure. Differently, our BH approach exploits a direct differential Joule heating below the nanostructure position and bypasses the conductive bottleneck represented by the oxide between the heater and the substrate. {\rodcolor As a consequence, our heating scheme largely outperforms TH performance reported so far in the literature and opens new possibilities for the investigation of TE effects on individual nanoscale active elements. In addition,} a careful design of the current-injection electrodes allows us to obtain a strong thermal gradient and to retain a controllable gating despite the presence of an electric field in the substrate. It is also crucial to note that in both cases heaters can be fabricated in any position and orientation on a Si/SiO$_2$ substrate and they are thus {\rodcolor applicable to the investigation of the effect of a thermal bias on nanostructures which are transferred by drop casting or similar methods, i.e. typically with a random position and orientation.}

The article is organized as follows: in Sec.~I we describe the fabrication procedure and discuss our BH design, also based on finite-element simulations; in Sec.~II we describe the operation of the heater and report its performance using resistive thermometers patterned on top of the Si/SiO$_2$ substrate; in Sec.~III we discuss heater performance and compare an experimental Raman mapping of the SiO$_2$ temperature with a detailed simulation. {\rodcolor In Sec.IV, the operation of the nanoheater is demonstrated on a single-nanowire field-effect transistor (FET) with a channel length of $1\,{\rm \mu m}$.}

\section{Heater design and fabrication}

\begin{figure}[b!]
\begin{center}
\includegraphics[width=0.48\textwidth]{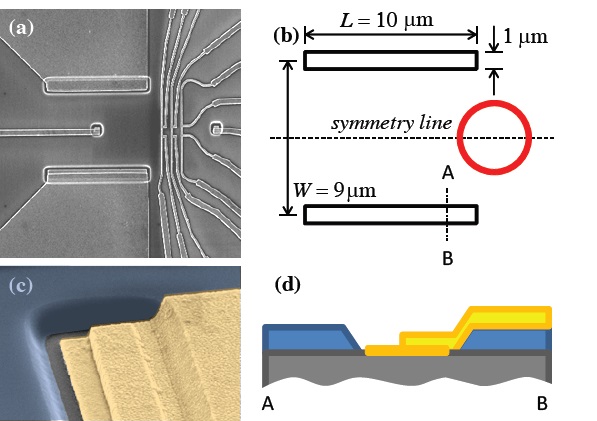}
\caption{\rodcolor (a) Scanning electron micrograph of one of the fabricated buried heaters. (b) Sketch of the contact geometry: the heater is designed to achieve the best differential heating conditions in the red circle are, which is the one where the nanostructure is supposed to reside. (c) Blow-up image of one of the current injection contacts used to feed current to the substrate through a hole in the $280\,{\rm nm}$-oxide covering the Si substrate (sample was tilted by $60$ degrees) . (d) Cross-section sketch of the contact region.}
\label{fig:SemDevice}
\end{center}
\end{figure}

Heaters were built starting from a highly-conductive ($\rho=0.001-0.005\,{\rm \Omega cm}$) Si substrate coated with a $280\,{\rm nm}$-thick oxide layer. The injection electrodes consist of two $1\times10\,{\rm \mu m^2}$ windows in the oxide, separated by $9\,{\rm \mu m}$. These were defined by e-beam lithography using PMMA resist and etched in a buffer-oxide etch solution for $5\,{\rm min}$. Immediately after etching a Ni/Au bilayer ($10/25\,{\rm nm}$) was deposited to contact the Si substrate. These contacts were connected to the large bonding pads using a wide and thicker Ti/Au ($10/100\,{\rm nm}$) evaporation step. It is important to note that the heater can be freely placed anywhere on the Si/SiO$_2$/ substrate and thus be aligned, for instance, to an existing nanostructure deposited on the oxide surface. However, the need to preserve the integrity of the nanostructures poses non-trivial constraints on the processing steps; for example, standard rapid thermal annealing of the heater contacts is not always possible. The device structure can be seen in the scanning electron micrograph and sketch of Fig. 2a and 2b. Further details of the contact region are visible in the tilted image of Fig. 2c and in the cross-section sketch of Fig. 2d. As visible in Fig. 2a, sets of four-wire resistive thermometers were fabricated together with the Ti/Au connection layer and later used to monitor the local temperature on the surface of the substrate.

\begin{figure*}[t]
\begin{center}
\includegraphics[width=0.96\textwidth]{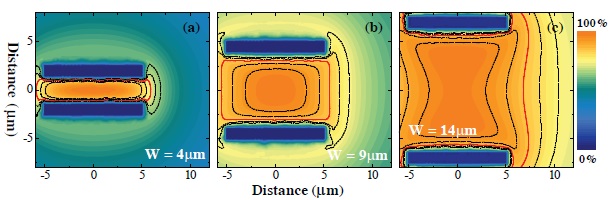}
\caption{\rodcolor Finite element estimate of the temperature profiles for different values of the $L/W$ ratio in the studied geometry and assuming an ideal thermal sinking at the heater contacts for sake of simplicity. The colorplot reports the temperature increase $\Delta T$ as a per cent of the maximum value in the heated regions. For a fixed $L=10\,{\rm \mu m}$ the (a), (b) and (c) panels report the case $W=4$, $9$ and $14\,{\rm \mu m}$, respectively. The red contour lines corresponds to the highest gradient along the symmetry line of the heater while the dashed lines are spaced by $5\%$ of the top $\Delta T$. An optimal uniformity at the top gradient position is achieved for $H\approx L$.}
\label{fig:Simulation}
\end{center}
\end{figure*}

The contact geometry visible in Fig. 2b was designed in order to achieve: (i) a large and uniform thermal bias at a position which is relatively far from the injection contacts; (ii) minimal non-uniformity in the electrostatic potential below the nanostructure under study in the presence of a heating current. A two-contact design was chosen because it allows to achieve a virtually vanishing voltage drop on the symmetry plane between the injection electrodes $H\pm$ when an asymmetric heater bias $V_{H\pm}=\pm V_H$ is applied. The ratio between the contact size $L$ and respective distance $W$ was decided based on finite-element simulations {\rodcolor indicating that the most uniform gradient is obtained for $W\approx L$. Calculation parameters and boundary conditions are discussed in further detail in the Supplementary Information while results are sketched in Fig.~\ref{fig:Simulation}}. The overall {\rodcolor size of the heater was decided based on the length of the studied nanowire structures, which is of the order of one to few microns.} Particular care was also taken in placing thermometer leads approximately along the expected heater constant-temperature lines, in order to avoid unwanted heat flow along the leads' and thermometers' electrodes.

\section{Heater operation} %%%%%%%%%%%%%%%%%%%%%%%%%%%%%%
% SELF ANNEALING 

\begin{figure}[h!]
\begin{center}
\includegraphics[width=.43\textwidth]{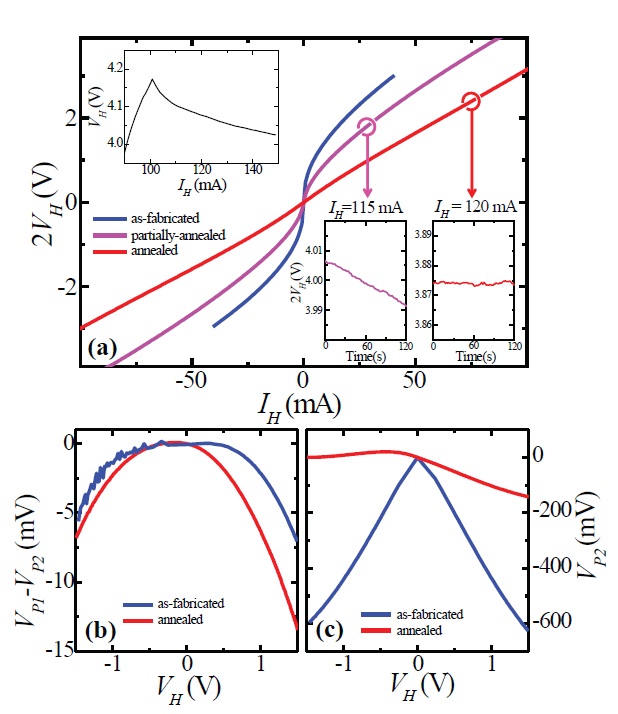}
\caption{(a) IV characteristics of the heater before (blue), during (magenta) and after (red) the self-annealing procedure. Top inset: a high driving current can lead to a time-dependent evolution of the IV. Bottom insets: the drift of the heater voltage $V_H$ at fixed $I_H$ and as a function of time. (b,c) Voltage difference  between the two substrate probes $V_{P1}$-$V_{P2}$ and individual value of $V_{P2}$. {\rodcolor Both panels report the characteristics ad various stages of the self-annealing procedure, according the the plot color.}}
\label{fig:HeaterCharacteristics}  
\end{center}
\end{figure}

The as-fabricated BH typically displays highly non-linear transport characteristics, as visible from curves in Fig.~\ref{fig:HeaterCharacteristics}a. This is not surprising since the Ni/Au contacts on Si are known to display a Schottky behavior {\rodcolor when no thermal treatment is performed and the semiconductor is not very strongly doped.} Such {\rodcolor non-ideal} behavior of the BH contacts can be detrimental since it leads to an uneven heating {\rodcolor caused by the voltage-drop occurring} at the contact position. In addition, non-ohmic contacts can cause a significant deviation from device symmetry under bias and, as a consequence, the back-gate voltage at the nanostructure position can become difficult to predict and control. Since {\rodcolor the heater} is meant to {\rodcolor be fabricated on a SiO$_2$/Si substrate already hosting the nanostructures to be studied}, we developed a procedure for ohmic-contact formation which avoids possibly dangerous standard annealing protocols. Indeed, a {\em self-annealing} procedure was performed exploiting the fact that a significant part of the injected power will be dissipated at the reverse-biased Schottky contact. The blue curve in Fig.~\ref{fig:HeaterCharacteristics}a represents an example of a typical IV curve of a BH before the self-annealing step. The heater was first biased using a current $I_H$, corresponding to a voltage drop defined as $2V_H$, consistently with the asymmetric voltage bias procedure that we will use in the following part of our analysis and with the notation of our previous work~\cite{Roddaro13}. Contact quality was improved by successive back and forth sweeps of the BH current $I_H$ between $-I_0$ and $+I_0$ for increasing values of $I_0$. The procedure was performed at room temperature. As visible in the top-left inset of Fig.~\ref{fig:HeaterCharacteristics}a, when sufficiently high currents are driven into the BH, heat dissipation at the reverse-biased contact starts to modify the contact properties and the IV curve displays a strong non-linear evolution due to its time-dependent shift to lower resistance values. This leads to a drop in the heater voltage $2V_H$ even while $I_H$ values keep increasing. It is important to stress that the process was repeated for both current directions in order to achieve good annealing for both contacts. The sweeps were repeated until no significant evolution was observed for a given $I_0$ value. Once the IVs became stable, $I_0$ was stepped to higher values. The process was repeated in small steps in order to avoid damaging the heater contacts and up to a top current $I_0=150\,{\rm mA}$. The pink curve in Fig.~\ref{fig:HeaterCharacteristics}a results from a partial self-annealing while the red curve corresponds to a fully self-annealed BH. The corresponding time drifts of the heater voltage $2V_H$ as a function of time and at a fixed bias current $I_H$ are shown in the bottom-right inset, with matching colors. In the case of the fully self-annealed BH, no drift is observed even at a significant current bias of $120\,{\rm mA}$.

As mentioned before, the Si substrate of the BH can also be used as a backgate to control the nanostructure properties by field effect. This can be achieved by applying a bias voltage $V_{H\pm}=V_{bg}\pm V_{H}$ to the two BH leads. In a perfectly symmetric device this would lead to a gate voltage $V_{bg}$ at the nanostructure position. However, given the residual asymmetry of the heater contacts, a direct measurement of the Si-bias value at the nanostructure position was carried out.  Since it is not possible to contact the Si directly below the nanostructure, the Si potential was probed at the two ends of the nanostructure ($V_{P1}$ and $V_{P2}$, see Fig.~\ref{fig:SemDevice}a). The values measured at the two probes are typically very consistent, with small deviations of few millivolts even at the highest $V_H$ values and both before and after the self-annealing procedure (see Fig.~\ref{fig:HeaterCharacteristics}b). Such a small difference indicates that biasing is uniform along the symmetry line of the BH, in the nanostructure region. The importance of the self-annealing procedure is also evident from curves in Fig.~\ref{fig:HeaterCharacteristics}c, where we report the absolute value of $V_{P1}$ for $V_{bg}=0$ and as a function of $V_H$. Before the annealing (blue curve), $V_{P1}$ can be a significant fraction of the applied $V_H$, consistently with the presence of two Schottky contacts. The situation is much improved after the annealing process (red curve) and the observed values are much closer to the ideal case $V_{P1}=V_{bg}=0$.

\section{Thermal performance}

\begin{figure}[h!]
\begin{center}
\includegraphics[width=0.48\textwidth]{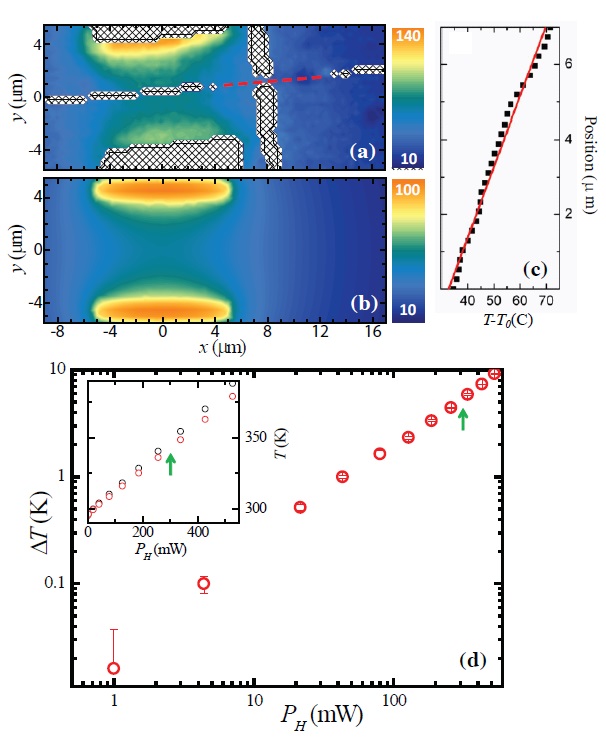}
\caption{\rodcolor (a) Raman map of the local temperature increase of the SiO$_2$/Si substrate in the vicinity of the heater and of the device measurement site. (b) Numerical simulation for a injected power of about $300\,{\rm mW}$ including the lead finite heat conductivity and contact dissipation. (c) The measured temperature profile along the red dashed line on the Raman map. (d) Temperature difference between two resistive thermometers spaced $1\,{\rm \mu m}$ (see Fig.~\ref{fig:SemDevice}a) as a function of the electrical power fed heater. The green arrows highlight the power used for the Raman map.}
\label{fig:Raman}
\end{center}
\end{figure}

{\rodcolor Figure~\ref{fig:Raman} illustrates the thermal characteristics of the BH in operation.} The two-dimensional heating profile was directly mapped by means of a micro-Raman technique which exploits the temperature dependence~\cite{raman83} of the Raman shift of the Si band at $520\,{\rm cm}^{-1}$. The sample was mounted on a three-axis piezo translator allowing precise positioning and electrical access to the heater. The sample was illuminated through a 100X objective with a numerical aperture of 0.7 with the 488-nm line of an Argon laser. The laser power on the sample was kept low ($1 {\rm mW}$) to avoid heating effects. Spectra were obtained by dispersing the scattered light with a 550-mm-focal-length spectrograph and focused onto a thermoelectrically cooled CCD. A full spectrum was acquired for each pixel of the map and post-processed by fitting the main Si peak with a Lorentzian curve to determine the precise position ($\omega$) of the band. The corresponding temperature was calculated as $T = T_0 - (\omega - \omega_0) \times 0.022\,{\rm cm^{-1}/^o C}$, where $T_0$ is the temperature of the device without any applied power (measured with a thermometer in close proximity to the $\mu$-Raman system) and $\omega_0$ the Raman shift of Si at $T_0$. 

{\rodcolor The heating profile $T-T_0$ is shown in Fig.~\ref{fig:Raman}(a)} for a biasing condition $V_H = 1.6\,{\rm V}$ and $I_H = 95\,{\rm mA}$, corresponding to a total injected power $P_H = 304\,{\rm mW}$. Hatched areas correspond to metal-covered areas of the device where the Si Raman signal could not be detected. {\rodcolor The measured profile indicates that contacts get significantly hotter with respect to the ideal case shown in Fig.~\ref{fig:Simulation}, indicating that an important effect is played by the non-perfect thermal sinking of the injection contacts as well as, most probably, by the local resistive heating at the metal-semiconductor interface. Figure~\ref{fig:Raman}(b) reports the result of a more refined numerical simulation including contact leads and an interface resistance at the electrode-Si contact. The resulting heating profile was calculated for a power dissipation of about $300\,{\rm mW}$ within the substrate and the heater contacts. A closer match to the actually observed $T$ profile is obtained and the overall heater behavior matches our expectations. In particular, the observed behavior confirms that a large and uniform gradient can be obtained with the chosen contact geometry. Further details about the numerical calculations are reported in the Supplementary material.} A profile of the experimental temperature along the white dashed line shown in Fig.~\ref{fig:Raman}a is reported in Fig.~\ref{fig:Raman}c. The gradient was determined to correspond to $5.3 \pm 0.2\,{\rm K}/\mu{\rm m}$ by linearly fitting the data (red line in the plot). 

{\rodcolor The effective BH performance was also cross-checked using a set of resistive thermometers,} as also visible in the right side of the heater in Fig.~\ref{fig:SemDevice}a: {\rodcolor these measurements were found to be consistent with Raman data}. In Fig.~\ref{fig:Raman}d we report the temperature difference between two thermometers spaced by $1\,{\rm \mu m}$. The temperature gradient was found to grow linearly with the power applied to the heater. In the device shown, a difference of almost $10\,{\rm K}$ was obtained by feeding about $500\,{\rm mW}$ into the BH circuit. Correspondingly, the absolute temperature of the thermometers - and thus that of the nanostructure - increased on average by less than $100\,{\rm ^oC}$  (see inset). The green arrows highlight the power setting used during the measurement of the Raman map of Fig.~\ref{fig:Raman}a. Linear interpolation of the available data yields $\Delta T(P_H=304\,{\rm mW})=5.29 {\rm K}$, and thermometer temperatures $T_1=347.98$~K and $T_2=342.69$~K. {\rodcolor These absolute values are consistent with the data that can be deduced from the Raman map.}

\section{Thermovoltage on single nanostructures}

{\rodcolor The above-described differential heating architecture was directly tested on a relatively-short InAs-nanowire FET with a channel length of $1\,{\rm \mu m}$, as visible in the scanning electron micrograph of Fig.~\ref{fig:nanowire}a. The device was fabricated starting by drop-casting n-doped InAs nanowires on top of the SiO$_2$/Si substrate pre-patterned with a set of markers. The nanowire position was then determined with respect to the markers. Heater and contacts were fabricated aligned to the randomly-deposited nanostructure. The contact leads of the nanowire were designed to work also as resistive thermometers and were obtained by a Ti/Au evaporation ($10/100\,{\rm nm}$). Prior to the contact deposition, wires were passivated by an ammonium sulfide solution in order to remove the native oxide on the InAs surface and promote contact formation. A blow-up of the device active region is visible in Fig.~\ref{fig:nanowire}b: note that it is located in front of the heater structure.

\begin{figure}[h!]
\begin{center}
\includegraphics[width=.40\textwidth]{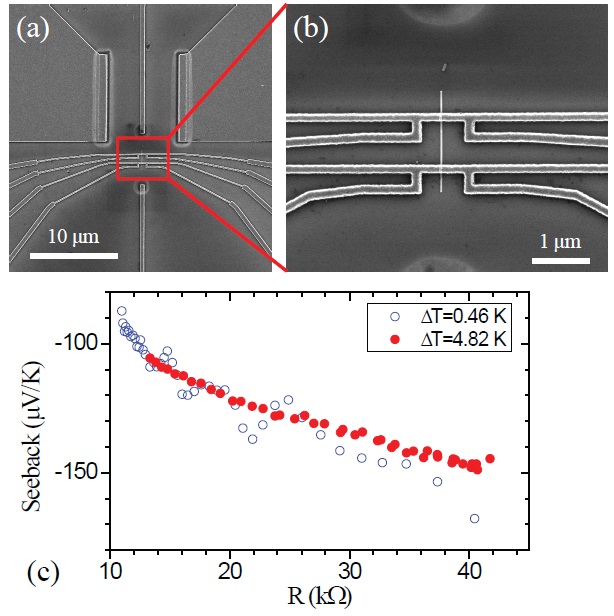}
\caption{\rodcolor (a) Scanning electron micrograph of the studied nanowire device and (b) blow-up of the isolated nanostructure. (c) Measurement of the field-effect evolution of the Seebeck coefficient plotted parametrically against the nanowire resistance. The application of a strong gradient of $\approx 5\,{\rm K/\mu m}$ allows to obtain a much clearer evidence of the effect.}
\label{fig:nanowire}
\end{center}
\end{figure}

The application of controlled, strong thermal gradients is crucial for the investigation of thermoelectric effects on single micrometric or even submicrometric nanostructured materials. Figure~\ref{fig:nanowire}c shows the result of a set of measurements of the Seebeck coefficient of single nanowires as a function of the field-effect-controlled carrier density. The resistance of the wire changes with the carrier density and is used as the abscissa in the parametric plot. Measurements were performed at room temperature. The advantage of the large gradient is evident by comparing the two curves obtained for a thermal bias $\Delta T$ of $0.46\,{\rm K}$ and $4.82\,{\rm K}$. While the two data are consistent, the large $\Delta T$ dataset is much cleaner since $S$ is calculated as the ratio of the thermovoltage and $\Delta T$ and its error obviously scales as $1/\Delta T$ for a given measurement protocol.}

%% Conclusions
\section{Conclusions}

We demonstrated a BH architecture which makes it possible to {\rodcolor reach} a thermal gradient {\rodcolor of almost $10\,{\rm K/\mu m}$} with an overall temperature increase of few tens of degrees with respect to the bath temperature {\rodcolor at the nanostructure position}. This result was obtained using the substrate as the active heating element and thus exploiting non-uniform Joule heating effects and bypassing the thermal impedance of the oxide layer. This scheme outperforms results reported in the literature and based on a TH approach: these are typically limited to gradients of fractions of ${\rm K/ \mu m}$\cite{Small03,Hochbaum09,Tian12}. {\rodcolor The differential heating architecture presented here was demonstrated on a single-nanowire device with an active channel of $1\,{\rm \mu m}$.} The possibility to achieve a gradient of this {\rodcolor magnitude} can significantly enhance the visibility of thermoelectric effects on the micrometer scale and reduce measurement errors, {\rodcolor enabling the investigation of even smaller nanometric active elements.}

The work was partly supported by the Marie Curie Initial Training Action (ITN)
Q-NET 264034 and by MIUR through the PRIN2009 project ``Dispositivi ad effetto di campo basati su nanofili e superconduttori
ad alta temperatura critica''.

%%%%%%%%%%%%%%%%%%%%%%%%%%%
%%% biblio
%%%%%%%%%%%%%%%%%%%%%%%%%%%


\begin{thebibliography}{99}

{\rodcolor
% ``Thermoelectric Cooling and Power Generation''
\bibitem{DiSalvo99} F. J. DiSalvo, {\em Thermoelectric Cooling and Power Generation}, Science {\bf 285}, 703 (1999).

% ``Cooling, Heating, Generating Power, and Recovering Waste Heat with Thermoelectric Systems''
\bibitem{Bell08} L. E. Bell, {\em Cooling, Heating, Generating Power, and Recovering Waste Heat with Thermoelectric Systems}, Science {\bf 321}, 1457 (2008).

% ``Thermoelectric Materials, Phenomena, and Applications: A Bird's Eye View''
\bibitem{Subramanian06} T. M. Tritt, M. A. Subramanian, {\em Thermoelectric Materials, Phenomena, and Applications: A Bird's Eye View}, MRS Bulletin {\bf 31}, 188 (2006).
}

% ``Thermoelectrics: Basic Principles and New Materials Developments''
\bibitem{BookTE} G. S. Nolas, J. Sharp, H. J. Goldsmid, {\em Thermoelectrics: Basic Principles and New Materials Developments},  Springer New York, 2001.

% ``Complex thermoelectric materials''
\bibitem{Snyder08} G.~J. Snyder, and E.~S. Toberer, {\em Complex thermoelectric materials}, Nature Mat. {\bf 7}, 105 (2008).

% ``Thermoelectricity in Semiconductor Nanostructures
\bibitem{Majumdar04} A. Majumbdar, {\em Thermoelectricity in Semiconductor Nanostructures}, Science {\bf 303}, 777 (2004).

% ``Nanostructured Thermoelectrics: Big Efficiency Gains from Small Features''
\bibitem{Vineis10} C.~J. Vineis, A. Shakouri, A. Majumdar, and M.~G. Kantzidis, {\em Nanostructured Thermoelectrics: Big Efficiency Gains from Small Features}, Adv. Mat. {\bf 22}, 3970 (2010).

% ``Thermal and Thermoelectric Transport in Nanostructures and Low-Dimensional Systems''
\bibitem{Shi12} L. Shi, {\em Thermal and Thermoelectric Transport in Nanostructures and Low-Dimensional Systems}, Nanoscale and Microscale Thermophysical Engineering {\bf 16}, 79 (2012).

% ``Silicon nanowires as effcient thermoelectric materials''
\bibitem{Boukai08} A.~I. Boukai, Y. Bunimovich, J. Tahir-Kheli, J.-K. Yu, W.~A. Goddard III, and J.~R. Heath, {\em Silicon nanowires as effcient thermoelectric materials}, Nature {\bf 451}, 06458 (2008).

% ``Thin-film thermoelectric devices with high room-temperature figures of merit''
\bibitem{Venkatasubramanian01} R. Venkatasubramanian, E. Siivola, T. Colpitts and B. O'Quinn, {\em Thin-film thermoelectric devices with high room-temperature figures of merit}, Nature {\bf 413}, 597 (2001).

% ``High-Thermeolectric Performance of Nanostructured Bismuth Antimony Telluride Bulk Alloys''
\bibitem{Poudel08} B. Poudel, Q. Hao, Y. Ma, Y. Lan, A. Minnich, B. Yu, X. Yan, D. Wang, A. Muto, D. Vashee, X. Chen, J. Liu, M.~S. Dresselhaus, G. Chen, and Z. Ren, {\em High-Thermeolectric Performance of Nanostructured Bismuth Antimony Telluride Bulk Alloys}, Science {\bf 320}, 634 (2008)

% ``Enhancement of Thermoelectric Efficiency in PbTe by Distortion of the Electronic Density of States''
\bibitem{Heremans08} J.~P. Heremans, V. Jovovic, E.~S. Toberer, A. Saramat, K. Kurosaki, A. Charoenphkdee, S. Yamanaka, and G.~S. Snyder, {\em Enhancement of Thermoelectric Efficiency in PbTe by Distortion of the Electronic Density of States}, Science {\bf 321}, 554 (2008).

% ``New Directions for Low-Dimensional Thermoelectric Materials''
\bibitem{Dresselhaus07} M.~S. Dresselhaus, G. Chen, M.~Y. Tang, R. Yang, H. Lee, D. Wang, Z. Ren, J.-P. Fleurial, and P. Gogna, {\em New Directions for Low-Dimensional Thermoelectric Materials}, Adv. Mat. {\bf 19}, 1043 (2007).

% ``High Thermoelectric Figure-of-Merit in Kondo Insulator Nanowires at Low Temperatures''
\bibitem{Zhang11} Y. Zhang, M.~S. Dresselhaus, Y. Shi, Z. Ren, and G. Chen, {\em High Thermoelectric Figure-of-Merit in Kondo Insulator Nanowires at Low Temperatures} Nano Lett. {\bf 11}, 1166 (2011).

% ``Large thermoelectric figure of merit in Si$_{1-x}$Ge$_x$ nanowires''
\bibitem{Shi10} L. Shi, D. Yao, G. Zhang, and B. Li, {\em Large thermoelectric figure of merit in Si$_{1-x}$Ge$_x$ nanowires}, Appl. Phys. Lett. {\bf 96}, 173108 (2010).

% ``Reversible Thermoelectric Nanomaterials''
\bibitem{Humphrey05} T. E. Humphrey, H. Linke, {\em Reversible Thermoelectric Nanomaterials}, Phys. Rev. Lett. {\bf 94}, 096601 (2005).

\bibitem{Svensson13} S. F. Svensson, E. A. Hoffmann, N. Nakpathomkun, P. M. Wu, H. Q. Xu, H. A. Nilsson, D. Sanchez, V. Kashcheyevs
and H. Linke {\em Nonlinear thermovoltage and thermocurrent in quantum dots}, New Jour. Phys. {\bf 15}, 105011 (2013).

\bibitem{Wu13} P. M. Wu, J. Gooth, X. Zianni, S. F. Svensson, J. G. Gluschke, K. A. Dick, C. Thelander, K. Nielsch and H. Linke, {\em Large Thermoelectric Power Factor Enhancement Observed in InAs Nanowires}, Nano Lett. {\bf 13}, 4080 (2013).

% ``Field-effect modulation of Seebeck coefficient in single PbSe nanowires''
\bibitem{Hochbaum09} W. Liang, A. I. Hochbaum, M. Fardy, O. Rabin, M. Zhang, P. Yang {\em Field-effect modulation of Seebeck coefficient in single PbSe nanowires}, Nano Lett. {\bf 9}, 1689 (2009).

{\rodcolor
% ``Manipulation of Electron Orbitals in Hard-Wall InAs/InP Nanowire Quantum Dots''
\bibitem{Roddaro11} S. Roddaro, A. Pescaglini, D. Ercolani, L. Sorba, and F. Beltram {\em Manipulation of Electron Orbitals in Hard-wall InAs/InP Nanowire Quantum Dots}, Nano Lett. {\bf 11}, 1695 (2011).

% ``Electrostatic Spin Control in InAs/InP Nanowire Quantum Dots''
\bibitem{Romeo12} L. Romeo, S. Roddaro, A. Pitanti, D. Ercolani, L. Sorba, and F. Beltram, {\em Electrostatic Spin Control in InAs/InP Nanowire Quantum Dots}, Nano Lett. {\bf 12}, 4490 (2012).
}

% ``Measuring Temperature Gradients over Nanometer Length Scales''
\bibitem{Hoffmann09} E.~A. Hoffmann, H.~A. Nilsson, J.~E. Matthews, N. Nakpathomkun, A.~I. Persson, L. Samuelson, and H. Linke, {\em Measuring Temperature Gradients over Nanometer Length Scales}, Nano Lett. {\bf 9}, 779 (2009).

% ``On errors in thermal conductivity measurements of suspended and supported nanowires using micro-thermometer devices from low to high temperatures''
\bibitem{Moore2011} Arden L Moore and Li Shi, {\em On errors in thermal conductivity measurements of suspended and supported nanowires using micro-thermometer devices from low to high temperatures}, Meas. Sci. Technol.  {\bf 22}, 015103 (2011).

% ``Thermal resistance of a nanoscale point contact to an indium arsenide nanowire''
\bibitem{Zhou2011}F. Zhou, A. Persson, L. Samuelson, H. Linke and Li Shi, {\em Thermal resistance of a nanoscale point contact to an indium arsenide nanowire}, Appl. Phys. Lett. {\bf 99} 063110 (2011).

% ``Modulation of Thermoelectric Power of Individual Carbon Nanotubes"
\bibitem{Small03}  J. P. Small, K. M. Perez, and P. Kim, {\em Modulation of Thermoelectric Power of Individual Carbon Nanotubes}, Phys. Rev. Lett. {\bf 91}, 256801 (2003).

% ``Thermopower measurement of individual single walled carbon nanotubes''
\bibitem{Small04} J. Small, {\em Thermopower measurement of individual single walled carbon nanotubes}, Microscale Thermophys. Eng. {\bf 8}, 1 (2004).

% "One-Dimensional Quantum Confinement Effect Modulated Thermoelectric Properties in InAs Nanowires"
\bibitem{Tian12} Y. Tian, M. R. Sakr, J. M. Kinder, D. Liang, M. J. MacDonald, R. L. J. Qiu, H.-J. Gao, and X. P. A. Gao, {\em One-Dimensional Quantum Confinement Effect Modulated Thermoelectric Properties in InAs Nanowires}, Nano Lett. {\bf 12}, 6492  (2012).

% "Giant Thermovoltage in Single InAs Nanowire Field-Effect Transistors"
\bibitem{Roddaro13} S. Roddaro, D. Ercolani, M. A. Safeen, S. Suomalainen, F. Rossella, F. Giazotto, L. Sorba, and F. Beltram, {\em Giant Thermovoltage in Single InAs Nanowire Field-Effect Transistors}, Nano Lett. {\bf 2013}, 3638 (2013).

% ``Anharmonic effects in light scattering due to optical phonons in silicon''
\bibitem{raman83} M. Balkanski, R. F. Wallis, and E. Haro, {\em Anharmonic effects in light scattering due to optical phonons in silicon}, Phys. Rev. B {\bf 28}, 1928 (1983).

% "Observation of Thermopower Oscillations in the Coulomb Blockade Regime in a Semiconducting Carbon Nanotube"
\bibitem{Llaguno04} M. C. Llaguno, J. E. Fischer, and A. T. Johnson, Jr., {\em Observation of Thermopower Oscillations in the Coulomb Blockade Regime in a Semiconducting Carbon Nanotube}, Nano Lett. {\bf 4}, 45 (2004).


\end{thebibliography}
\end{document}